\documentclass[a4paper,11pt]{article}
\usepackage{jcappub} 
\usepackage{lineno}
\usepackage{graphicx}%
\usepackage{multirow}%
\usepackage{amsmath,amssymb,amsfonts}%
\usepackage{amsthm}%
\usepackage{mathrsfs}%
\usepackage[title]{appendix}%
\usepackage{xcolor}%
\usepackage{textcomp}%
\usepackage{manyfoot}%
\usepackage{booktabs}%
\usepackage{algorithm}%
\usepackage{algorithmicx}%
\usepackage{algpseudocode}%
\usepackage{listings}%
\usepackage{breakurl}%

\usepackage{bm}
\usepackage{subcaption}
\usepackage{float}
\usepackage{bbm}
\usepackage[free-standing-units=true]{siunitx}
\usepackage{comment}
\usepackage{dsfont}

\title{\boldmath Information Field Theory with JAX infers Air Shower Electric Currents from Antenna Signal Traces}

\author[a]{Maximilian Straub}
\author[b, c]{, Torsten Enßlin}
\author[a]{, Martin Erdmann}
\author[b,d]{, Philipp Frank}
\author[a]{, Stefanie Girod}
\author[a]{and Mike Zingler}
\affiliation[a]{RWTH Aachen University, Physics Institute 3A, \\Otto-Blumenthal-Str., 52074 Aachen, Germany}
\affiliation[b]{Max Planck Institute for Astrophysics,\\Karl-Schwarzschild-Str. 1, 85748 Garching, Germany}
\affiliation[c]{Deutsches Zentrum für Astrophysik, \\Postplatz 1, 02826 Görlitz, Germany}
\affiliation[d]{Kavli Institute for Particle Astrophysics \& Cosmology (KIPAC), Stanford University,\\ CA 94305, Stanford, USA}

\emailAdd{erdmann@physik.rwth-aachen.de}

\abstract{Direct imaging of cosmic-ray-induced particle showers during daylight is a long-standing challenge in astroparticle physics. A promising avenue for capturing images of these showers is through the radio emissions generated by their electrically charged particles. Their corresponding current vectors evolve as the particle shower propagates through the Earth's atmosphere, leading to a characteristic time-dependent electric field in an antenna array. In this work, we harness modern Bayesian inference techniques within the Python toolkit for numerical information field theory NIFTy, coupled with the high-performance numerical computing capabilities of the Python library JAX. This combination enables us to reconstruct detailed information on the particle shower and its development from data collected by a ground-based antenna array. The source model contains a vector-valued current sheet in the shower plane, a finite shower-front thickness behind the leading front, and a differentiable radio-response model based on retarded-time propagation. The finite thickness is required to reproduce the temporal structure of the measured electric-field traces, especially for geometries affected by Cherenkov compression. For the first time, both the longitudinal and transverse geomagnetic components of the current vectors across the shower plane are derived during the shower development, as demonstrated using simulated air showers. Our approach opens a pathway for detailed imaging of cosmic-ray showers, potentially advancing our understanding of high-energy astrophysical processes.}

\begin{document}
\maketitle
\flushbottom

\section{Introduction}

Following classical electrodynamics, we can reconstruct the direction and magnitude of an electric current by measuring the resulting electromagnetic field \cite{jackson_classical_1999}. Scaling this situation up to an interplay of multiple temporally and spatially variable currents is highly non-trivial and subject of current research. 

A specific research question emerges when examining particle showers in the Earth’s atmosphere and their associated electromagnetic radiation. These showers are induced by the collision between high-energy cosmic particles and molecules in the Earth's atmosphere. A single particle shower starts high in the atmosphere and approaches the Earth at almost the speed of light, centering along the original cosmic particle direction. Transverse to that direction, the secondary particles approximately form a disk. Initially, this disk widens until the energy of the cosmic particle is distributed among billions of particles, most of which are then absorbed in the atmosphere. 

In this work, we focus on both temporally and spatially variable electric currents caused by the electrically charged particles of such a particle shower. They arise due to deflection in the Earth's magnetic field and the separation of positive and negative charge carriers. Overall, this results in short-term electromagnetic waves in the radio range of a few 10-100 MHz, which can be detected using antenna arrays \cite{Escudie:2019tlt, LOPES:2021ipp, PierreAuger:2012ker, PierreAuger:2016vya, PierreAuger:2023lkx, Thoudam:2015lba}. These waves propagate to earth essentially undisturbed.

Various detailed simulation programs have been developed in the past decades that describe radiation in the radio frequency range from air showers very well \cite{wernerMacroscopicTreatmentRadio2008,Alvarez-Muniz:2014wna, Alameddine:2024cyd}. These programs enable a forward simulation starting from the particles causing the radiation up to the electric field observation at radio antennas. For the reverse process, namely drawing conclusions from the antenna signals on the development of the particle shower, these simulation programs are computationally intensive which limits their use for inferences.

In this work, we investigate the principle possibilities of a temporal and spatial reconstruction of the particle shower evolution from the measured signals of an antenna array. We simplify the challenge to the following abstract situation. The shower is represented in a coordinate system aligned with its propagation axis. In this frame, a compact current-carrying structure moves at nearly the speed of light through a large atmospheric volume with a one-dimensional refractive-index gradient. 

The source is generated as a vector-valued current sheet over the transverse shower plane and along the shower development. This sheet is then redistributed along a finite thickness coordinate behind the leading shower front. The resulting semi-parametric model combines flexible correlated fields for the local current structure with physically motivated envelopes for the longitudinal and lateral extent and with a radial model for the shower-front thickness.

We combine two modern technologies that make the inference process for this simplified situation manageable. 
With the Numerical Information Field Theory (NIFTy) package \cite{Edenhofer2024} based on the information field theory (IFT) \cite{ensslinInformationTheoryFields2019} we perform the vector field reconstruction of the current density based on Bayes probability. The JAX library \cite{jax2018github} enables us to model the evolving vector field in highly parallel computations on GPUs. 

This work is structured as follows: First, we outline the principle of antenna reciprocity in their specific application to particle showers. Next, we discuss our approach to the numerical calculations. Then we introduce the two key technologies employed in our study - IFT and JAX. Finally, we formulate the shower model that needs to be adapted based on the observational data. We then describe the inference procedure and present benchmarks on the reconstruction performance using simulated air showers. Finally, we provide a summary of our findings.

\section{Lorentz Reciprocity Theorem}

In this section, we first recall the principle of antenna reciprocity and how to use it to obtain a Green's function to simplify the calculation of radio emissions. 
\begin{figure}[h!]
    \centering
    \includegraphics[width=0.34\linewidth]{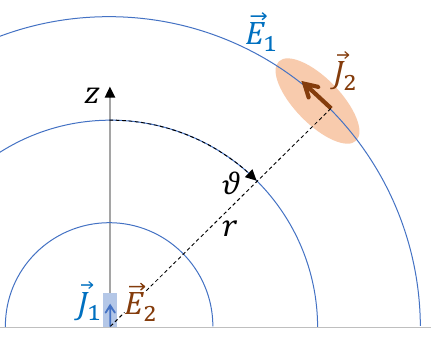}
    \caption{Lorentz reciprocity theorem applied to an air shower and a dipole antenna}
    \label{fig:Lorentz}
\end{figure}

The Lorentz reciprocity theorem describes the interconnection between two current densities $\vec{J}_i$ and their electric fields $\vec{E}_i$ (Fig.\ref{fig:Lorentz}), here denoted in temporal Fourier space using the hat symbol:
\begin{align}
    \int{\vec{\hat{E}}_2(\vec{x},\omega)\,\vec{\hat{J}}_1(\vec{x},\omega)}\,dV
    &=\int{\vec{\hat{E}}_1(\vec{x},\omega)\,\vec{\hat{J}}_2(\vec{x},\omega)}\,dV
    \label{eq:lorentz}
\end{align}
The vector $\vec{x}$ refers to a location in coordinate space, the frequency $\omega$ here indicates a Fourier transformation from the time domain, and $\int dV$ denotes the volume integral. Applying this theorem to the situation of an antenna for measurements of radio emission from air showers, we present here the key arguments only. The corresponding exact mathematical calculations are quite involved and can be found in Ref.~\cite{Riegler:2020tzx}.

The time-varying current density $\vec{J}_2$ produced by an air shower generates a time-varying electric field $\vec{E}_2$ in the antenna, which is measured there as an induced voltage $U_{\text{ind}}$. Conversely, a current density $\vec{J}_1$ in the antenna generates a field with electric field vector $\vec{E}_1$ in the atmosphere. At first glance, these appear to be two independent processes. 

However, recall the transmitting and receiving properties of antennas are reciprocal: A current density in the antenna generates an electromagnetic signal in the atmosphere. Conversely, the same current density can be induced by reception of an electromagnetic signal originating from the atmosphere. Thus, when a short pulse is injected into the antenna and the resulting electromagnetic wave propagates through the atmosphere, the corresponding propagation effects required to control the reception of electromagnetic waves from the air shower are encoded. 
As part of antenna reciprocity, it is also necessary to ensure that the emission of the air shower current density $\vec{J}_2$ lies within the space-time domain in which the Green’s function explained below is valid.

To take advantage of the theorem, let us first consider the antenna as an infinitesimally small dipole of length $ds$, oriented perpendicular to the Earth's surface ($z$-di\-rec\-tion). 
We now use Lorentz reciprocity to establish an interconnection between the processes
of signal reception and signal transmission at the antenna, as seen in Fig.~\ref{fig:Lorentz}.
For the process of signal transmission, we formally assign a delta-shaped current pulse with charge
$Q$ to the current density $\vec{J}_1$ on the left hand side of (\ref{eq:lorentz}), here denoted in the time domain:
\begin{align}
\vec{J}_1 = \delta(t)\,\delta(\vec{x})\,Q\,ds\,\vec{e}_z     
\end{align}
Given that $\vec{J}_1$ is a current density, the product $Q\,\delta(\vec{x})$ yields the dimension of a charge density, while $ds\,\delta(t)$ ensures the dimension of velocity.
The choice of the delta function allows the integral on the left-hand side of (\ref{eq:lorentz}) to be solved. 
Consequently, the electric field component $\vec{E}_2 \, \vec{e}_z=E_{2 z}$ at the position of the antenna, originating from the air shower current density $\vec{J}_2$, can be determined. Thus the left-hand side of (\ref{eq:lorentz}) represents the signal reception process at the detector and can be measured in practice as an induced voltage:
\begin{align}
     E_{2 z}\, ds = U_{\text{ind}}
\end{align}
On the right-hand side of (\ref{eq:lorentz}), the corresponding electric field response $\vec{E}_1$ of the delta pulse, which is often referred to as the weighting field $\vec{E}_1 = \vec{E}_{\text{w}}$, needs to be inserted. The analytical expression of this weighting field can be derived by superimposing periodic currents whose far-field solutions for the electric field contain well-known terms from electrodynamics \cite{kraus_1988}:
\begin{align}
    &\vec{E}_{\text{w}}(\vec{x},t)=Q\,ds\cdot\frac{n^2_{\text{eff}}\,\sin{\vartheta}}{4\pi\varepsilon\,c^2\,r} \,\delta^\prime\Big(t-\frac{n_{\text{eff}}\,r}{c}\Big)\,\vec{e}_\vartheta
\end{align}
Here $\vartheta$ refers to the zenith angle, $\varepsilon$ is permittivity of the medium, $c$ is the vacuum velocity of light, $r$ is the spatial distance between the air shower emission and the antenna, and the time derivative $\delta^\prime$ of the delta function includes the travel time $n_{\text{eff}}\,r/c$ from emission to observation. By $n_\text{eff}$ we describe an effective refractive index between the dipole and a point in space at distance $r$, which is obtained by an averaging integral along a straight line between the two points. Approximating the radiation trajectory as a straight line with a constant (average) refractive index works well in air ~\cite{wernerMacroscopicTreatmentRadio2008}. This is akin to treating the atmosphere as a homogeneous medium between two points. 

Thus, the Lorentz reciprocity theorem (\ref{eq:lorentz}) establishes here a relationship between the current densities $\vec{J}_2$ triggered by the air shower and the corresponding electric field component $E_{2 z}$ in an infinitesimal vertical oriented dipole antenna on earth:
\begin{align}
    E_{2 z}(t)\, ds  = \frac{1}{Q}\cdot\int\hspace*{-1mm} dV\int\limits_{-\infty}^{\infty}{\hspace*{-2mm}\vec{E}_{\text{w}}(\vec{x},t^\prime)\, \vec{J}_2(\vec{x},t-t^\prime)}\,dt^\prime
    \label{eq:E2z}
\end{align}
The time integral encodes the adaptation of the times of the weighting field $\vec{E}_{\text{w}}$ to the emission times of the air shower current density $\vec{J}_2$ (Fig.\ref{fig:Lorentz}).

To account for a real dipole antenna oriented in the $z$-direction, both sides of (\ref{eq:E2z}) must be convolved with the corresponding antenna response function $h$. On the left-hand side, this leads to the measured electric field $E_{2 z, h}$ as observed within the realistic antenna. On the right-hand side, $\vec{E}_{\text{w}}$ is also convolved with the antenna response $h$, resulting in the Green's function $\vec{K}$ of the dipole antenna, which needs to be determined only once:
\begin{align}
    \vec{K}(\vec{x},t) &= \frac{1}{Q}\,\int\limits_{-\infty}^{\infty}{\vec{E}_{\text{w}}(\vec{x},t^\prime)\, h(t-t^\prime)}\,dt^\prime\\
    &=\underbrace{ds\cdot\,\frac{n^2_{\text{eff}}}{4\pi\varepsilon\,c^2\,r} \,h^\prime\Big(t-\frac{n_{\text{eff}}\,r}{c}\Big)}_{\equiv K^\prime(\vec{x},t)}\,\sin{\vartheta}\,\vec{e}_\vartheta
    \label{eq:Greens}
\end{align}
The function $h^\prime$ is the time derivative of the antenna response characteristic.

Finally, we aim to determine the three-dimensional electric field of the air shower on the ground by measuring the electric field components as observed within three realistic dipole antennas oriented along the $x$-, $y$-, and $z$-directions, all located at the coordinate origin $\vec{0}$. The corresponding electric field vector can be described by a matrix-vector multiplication $\mathbf{D}\,\vec{J}_2$ where the zenith dependence $\sin{\vartheta}$ and $\vec{e}_\vartheta$ of $\vec{K}$ are absorbed into the matrix $\mathbf{D}$ and the scalar $K^\prime$ of the Green's function (\ref{eq:Greens}) remains:
\begin{align}
    \vec{E}_{2,h}(\vec{0},t)\,ds &=\int dV\,\int\limits_{-\infty}^{\infty}{K^\prime(\vec{x},t^\prime)\, \mathbf{D}\,\vec{J}_2(\vec{x},t-t^\prime)}\,dt^\prime
    \label{eq:inference}
\end{align}
The matrix denotes:
\begin{align}
\mathbf{D}=\scriptsize{\frac{1}{r^2}\left(\begin{array}{ccc}
-(y^2+z^2) & x y & x z \\
y x & -(z^2+x^2) & y z \\
z x & z y & -(x^2+y^2)
\end{array}\right)}
\end{align}
This matrix relates how sensitive each dipole is in the direction of the emission origin. Effectively, the convolution between Green's function $K^\prime\,\mathbf{D}$ and the current density $\vec{J}_2$ solves three tasks:
\begin{itemize}
    \item Direction-dependent weighing,
    \item Distance-dependent scaling,
    \item Distance-dependent shift in time.
\end{itemize}

Equation (\ref{eq:inference}) is our first milestone to describe a moving set of electric current densities along a trajectory in the atmosphere inducing an electromagnetic field in an antenna device on earth. Calculating the received electromagnetic signal in the three dipole antennas can be done with the described Green's function approach.

What we still need to do next is to carefully reproduce the time-shifting of the Green's function $K^\prime$, produced by the time-shifted argument $t-r\,n_\text{eff}/c$ inside the time derivative $h^\prime$ (\ref{eq:Greens}) of the response function.

We will later use the relationship (\ref{eq:inference}) to infer the space-time dependent current density profile in the air shower based on the electric fields $\vec{E}_{2,h}(\vec{x}_\text{obs},t)$ measured in an array of antennas, ultimately allowing us to draw conclusions about the temporal evolution of the air shower in the atmosphere.

\section{Numerical Calculation}
To accomplish the numerical calculations, we describe the particle shower in a coordinate system aligned with its propagation axis. The source is represented by a moving current sheet in the transverse shower plane, together with a finite thickness behind the leading shower front. The shower moves at almost the speed of light through a large volume with a one-dimensional, increasing refractive index, representing the Earth's atmosphere. Below we outline the main concepts; a comprehensive discussion of the numerical details can be found in \cite{straub2026}.

\paragraph{Shower-aligned coordinates:}

We use a shower-aligned orthonormal basis,
\begin{align}
    \vec{e}_1 &\equiv \vec{e}_{\vec{v}\times\vec{B}} \,,\\
    \vec{e}_2 &\equiv \vec{e}_{\vec{v}\times(\vec{v}\times\vec{B})} \,,\\
    \vec{e}_3 &\equiv \vec{e}_{\vec{v}} \,,
    \label{eq:shower_basis}
\end{align}
where $\vec{B}$ is the geomagnetic field and $\vec{v}$ denotes the shower direction.

\begin{figure}[h!]
    \centering
    \includegraphics[width=0.5\textwidth]{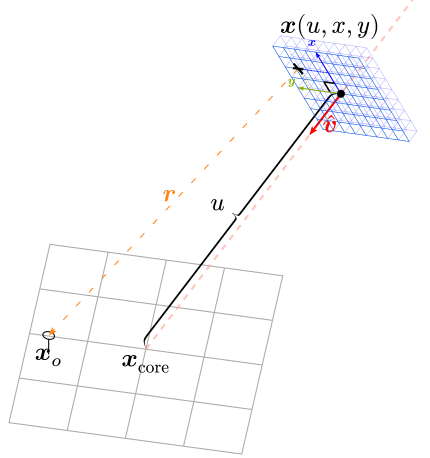}
\caption{Coordinate system used for the shower source. The leading shower front is described on a two-dimensional grid centered on the shower axis. The shower front moves along the $u$-axis, and the transverse coordinates are denoted by $x$ and $y$. The finite thickness is described by the coordinate $h$ behind the leading front. The $u$-axis is aligned with the shower velocity $\vec{v}$ and extends upward into the atmosphere from its point of intersection with the ground at $\vec{x}_{\textrm{core}}$. An observer is located at $\vec{x}_{\mathrm{obs}}$ on the ground.}    
    \label{fig:box}
\end{figure}

Let $\vec{e}_{\vec{v}}$ be the unit vector along the shower axis and $\vec{x}_{\mathrm{core}}$ the shower core position on the ground (Figure~\ref{fig:box}). A point on the shower axis is parameterized by the coordinate $u$,
\begin{align}
    \vec{x}(u) = \vec{x}_{\mathrm{core}} - u\,\vec{e}_{\vec{v}} .
    \label{eq:shower_axis}
\end{align}
The sign convention is chosen such that $u=0$ corresponds to the impact point on the ground and increasing $u$ corresponds to earlier stages of the shower development. The transverse shower plane is spanned by the two unit vectors with $\vec{e}_{\vec{v}\times\vec{B}}$ chosen along the geomagnetic direction $\vec{v}\times\vec{B}$ and $\vec{e}_{\vec{v}\times(\vec{v}\times\vec{B})}$ perpendicular to it in the shower plane.

The position of the leading shower front is therefore
\begin{align}
    \vec{x}_{\mathrm{front}}(u,x,y)
    =
    \vec{x}_{\mathrm{core}}
    -
    u\,\vec{e}_{\vec{v}}
    +
    x\,\vec{e}_{\vec{v}\times\vec{B}}
    +
    y\,\vec{e}_{\vec{v}\times(\vec{v}\times\vec{B})} .
    \label{eq:front_position}
\end{align}
The emission time is tied to the position of the moving shower front,
\begin{align}
    t_{\mathrm{emit}}(u) = -\frac{u}{c}.
    \label{eq:emission_time_u}
\end{align}
Thus, the temporal development of the shower is represented by its spatial evolution along the shower axis.

To account for the finite extension of the shower front, we introduce a thickness coordinate $h$. The value $h=0$ denotes the leading shower front, while $h>0$ denotes positions trailing behind it. A source point including this finite thickness is located at
\begin{align}
    \vec{x}(u,x,y,h)
    =
    \vec{x}_{\mathrm{front}}(u,x,y)
    -
    h\,\vec{e}_{\vec{v}} .
    \label{eq:thick_position}
\end{align}

\paragraph{Arrival time at the observer:}

\begin{figure}[h!]
    \centering
    \includegraphics[width=0.55\linewidth]{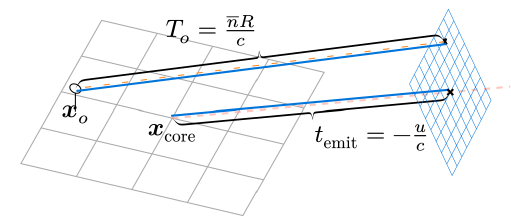}
\caption{Calculation of the arrival time. The source moves along the $u$-axis, defining its emission time. The light travel time from the source to the observer is obtained by calculating the optical path length along the straight line between source and observer. The arrival time is therefore the $u$-dependent emission time plus the light travel time depending on the full source coordinates $(u,x,y,h)$.}
    \label{fig:time_shift}
\end{figure}

When viewed from the Earth, the shower front moves with velocity close to $c$ along the shower axis towards the observer located at $\vec{x}_{\mathrm{obs}}^{\,a}$, where the upper index $a$ labels the antenna (Figure~\ref{fig:time_shift}). A signal emitted at time $t_{\mathrm{emit}}$ (\ref{eq:emission_time_u}) from the source point $\vec{x}(u,x,y,h)$ arrives at observer $a$ at the retarded time
\begin{align}
    \tau^a(u,x,y,h)
    =
    -\frac{u}{c}
    +
    \frac{
    n_{\mathrm{eff}}\!\left(\vec{x}_{\mathrm{obs}}^{\,a},\vec{x}(u,x,y,h)\right)
    }{c}
    \,
    \left|
    \vec{x}_{\mathrm{obs}}^{\,a}
    -
    \vec{x}(u,x,y,h)
    \right| .
    \label{eq:arrival_time_thick}
\end{align}
Here $n_{\mathrm{eff}}$ denotes the effective refractive index obtained by averaging the refractive index along the line of sight from the emission point to the observer. For radio emission in air, bending of the propagation path is negligible to good approximation, so that straight-line propagation can be used~\cite{wernerMacroscopicTreatmentRadio2008}.

For a sequence of emissions from a relativistically moving shower, the mapping from the shower-development coordinate $u$ to the observer time is non-trivial. A unit interval in $u$ can be compressed or stretched in observer time, and the mapping can even become locally non-monotonic. This behavior is the numerical manifestation of Cherenkov compression and must be treated carefully in the forward calculation of the generative model.

\paragraph{Retarded-time deposition:}

The Green's function derived in (\ref{eq:inference}) maps the current density $\vec{j}$ in the air shower to the electric field observed in the antenna. For the numerical implementation it is convenient to first write the accumulated pre-time-derivative field $\vec{A}^{\,a}$ for observer $a$ as
\begin{align}
    \vec{A}^{\,a}(t)
    =
    \frac{1}{c}
    \int du\,dx\,dy\,dh\,
    K^a\!\left(\vec{x}(u,x,y,h)\right)\,
    \vec{j}(u,x,y,h)\,
    \delta\!\left(t-\tau^a(u,x,y,h)\right).
    \label{eq:pre_derivative_field}
\end{align}
The kernel $K^a$ contains the distance scaling, the effective refractive index, and the transverse projection of the current relative to observer $a$. The measured electric field at this observer is obtained from the corresponding time derivative, $\vec{E}^{\,a}(t)=\partial\vec{A}^{\,a}(t)/\partial t$. In the implementation, this derivative is not evaluated by explicitly forming and differentiating $\vec{A}^{\,a}(t)$. Instead, the edge structure of the deposited signal is accumulated directly on the observer-time grid.

\begin{figure}[h!]
    \centering
    \includegraphics[width=0.55\linewidth]{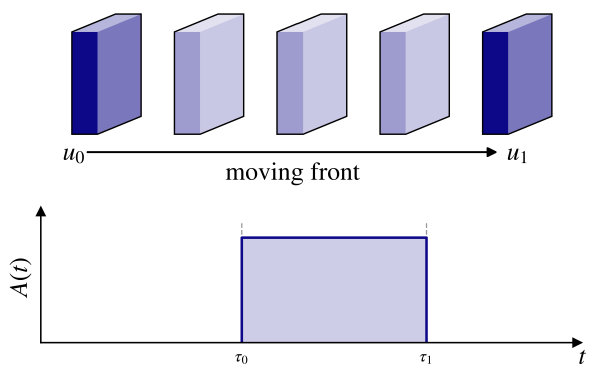}
    \caption{Deposited signal from an infinitely flat cell, traveling along a
straight line. Top: A thin cell moving along a straight line between two endpoints. The shading is only a visual aid and the front is treated as infinitely thin. The geometry dependent factors are assumed to be constant over the extent of the path. Bottom: The resulting signal is a box, whose extent stems solely from the arrival times of the path endpoints $u_0$ and $u_1$.}
    \label{fig:thin_retarded_time}
\end{figure}

The source is discretized into transverse cells of area $\Delta A=\Delta x\,\Delta y$, segments along the shower axis of length $\Delta u$, and thickness bins of width $\Delta h_i$. For a thin shower front, one moving source segment maps to a finite interval in observer time (Figure~\ref{fig:thin_retarded_time}). If the front arrival times at the segment endpoints are denoted by $\tau_0^a$ and $\tau_1^a$, the contribution of this segment is deposited over the support
\begin{align}
    t_{\mathrm{lo}}^a &= \min(\tau_0^a,\tau_1^a),\\
    t_{\mathrm{hi}}^a &= \max(\tau_0^a,\tau_1^a).
    \label{eq:thin_support}
\end{align}
The height of the deposited box is chosen such that the integrated contribution is conserved. This area-preserving deposition implements the Jacobian of the retarded-time map and naturally accounts for time compression.

\paragraph{Finite shower-front thickness:}

The finite shower-front thickness redistributes the current sheet along $h$. The current sheet in the shower plane is denoted by $\vec{S}(u,x,y)$, and the normalized thickness weights are denoted by $w_i(r)$, where $r=\sqrt{x^2+y^2}$ is the lateral distance from the shower axis. The piecewise constant current density in thickness bin $i$ is written as
\begin{align}
    \vec{j}_i(u,x,y)
    =
    \frac{\vec{S}(u,x,y)\,w_i(r)}{\Delta h_i},
    \qquad
    \sum_i w_i(r)=1 .
    \label{eq:current_density_from_sheet}
\end{align}
The normalization ensures that the thickness model redistributes the current behind the front without changing the integrated current sheet.

For the propagation calculation, the geometry-dependent factors of the kernel are evaluated at the leading front, while the thickness dependence of the arrival time is retained. To first order, the arrival time of a point behind the front is
\begin{align}
    \tau^a(h)
    \simeq
    \tau^a(0)
    +
    \alpha^a h,
    \qquad
    \alpha^a
    =
    \frac{n_{\mathrm{eff}}}{c}\,
    \hat{\vec{r}}^{\,a}\cdot\vec{e}_{\vec{v}},
    \label{eq:height_linearization}
\end{align}
where $\hat{\vec{r}}^{\,a}$ is the unit vector from the source point to observer $a$. The parameter $\alpha^a$ describes how a distance behind the front is converted into a delay or advance in observer time.

\begin{figure}[h!]
    \centering
    \includegraphics[width=0.6\linewidth]{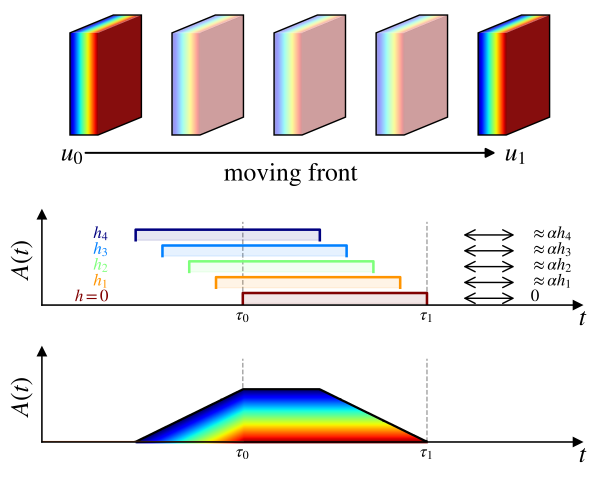}
    \caption{Deposited signal from a cell with finite extent, traveling along a straight line. The color marks the height $h$ behind the cell. Top: A cell moving between two endpoints. The thickness introduces an $h$-dependent time delay, which linearly expanded is $\propto\alpha\,h_i$. Middle: Decomposing the height into a number of thin cell placed at distance $h_i$ to the front, each would produce a box, shifted by the corresponding delay. Bottom: The resulting signal. Summing the contributions for all heights results in a trapezoid. The color in the trapezoid indicates where the height contributes. The support of the trapezoid is larger than that of the signal for the infinitely thin case.}
    \label{fig:thick_retarded_time}
\end{figure}

For a moving segment with finite thickness, as shown in Figure~\ref{fig:thick_retarded_time}, the time support is obtained from the four combinations of segment endpoints and thickness-bin boundaries,
\begin{align}
    t_{00}^a &= \tau_0^a + \alpha_0^a h_i, &
    t_{01}^a &= \tau_0^a + \alpha_0^a h_{i+1},\\
    t_{10}^a &= \tau_1^a + \alpha_1^a h_i, &
    t_{11}^a &= \tau_1^a + \alpha_1^a h_{i+1}.
    \label{eq:thickness_corners}
\end{align}
The deposited support is then approximated by
\begin{align}
    t_{\mathrm{lo},i}^a &= \min(t_{00}^a,t_{01}^a,t_{10}^a,t_{11}^a),\\
    t_{\mathrm{hi},i}^a &= \max(t_{00}^a,t_{01}^a,t_{10}^a,t_{11}^a).
    \label{eq:thick_support}
\end{align}
Under the local linearization, the exact contribution of such a segment is trapezoidal in observer time. In the renderer, this trapezoid is approximated by an equal-area box on the interval $[t_{\mathrm{lo},i}^a,t_{\mathrm{hi},i}^a]$. The box edges are deposited with sub-bin precision on the observer-time grid. This procedure preserves the integrated contribution and keeps the response differentiable and computationally efficient.

The total electric field at observer $a$ is obtained by accumulating all edge-deposited contributions on the observer-time grid:
\begin{align}
    \vec{E}^{\,a}(t)
    =
    \sum_{\mathrm{seg}}
    \sum_{\mathrm{cell}}
    \sum_i
    \Delta\vec{E}^{\,a}_{\mathrm{seg},\mathrm{cell},i}(t).
    \label{eq:electric_field}
\end{align}
The sums run over the shower-development segments, the transverse cells in the shower plane, and the thickness bins behind the leading shower front. The term $\Delta\vec{E}^{\,a}_{\mathrm{seg},\mathrm{cell},i}(t)$ denotes the edge-deposited electric-field contribution of the corresponding source element after applying the Green-function projection, distance scaling, and retarded-time mapping.

\section{Technologies}

Here we explain the technical basis with which we reconstruct the development of the particle shower in the atmosphere from its radio emission using antenna signals on ground.

Our challenge is high-dimensional because the shower current is represented on a source grid that resolves the shower development along the propagation axis, the transverse structure in the shower plane, and the finite shower-front thickness. The current values on this grid can jointly be understood as a physical field. Information Field Theory (IFT) provides a suitable mathematical basis for accommodating such a field and modifying it according to specified boundary conditions \cite{ensslinInformationTheoryFields2019}.

The source model is evaluated in a co-moving shower coordinate system, where the shower evolution is encoded along the discretized shower axis and the finite shower-front thickness is represented by an additional coordinate behind the leading front. The response operator must then propagate this current distribution to all antennas using the observer-dependent retarded-time map. A parallel calculation of these source elements is mandatory to keep the computing time manageable. The JAX library offers a highly suitable working environment for parallel operations on these field quantities \cite{jax2018github}.

\subsection{Information Field Theory}

A particular area of information theory is its application to physical fields. In the context of Information Field Theory (IFT) \cite{ensslinInformationTheoryFields2019,correlatedfield}, field-like quantities can be processed with Bayesian probabilities. In particular, IFT allows tackling research questions with an infinite number of degrees of freedom, i.e. continuous physical fields and their complementary discrete representations. With the help of prior physics knowledge about the question to be solved, probability-based assumptions can be introduced, thus reducing the number of degrees of freedom. This enables field expectation values to be calculated.

Given data $d$ and a model with parameter of interest $s$, Bayes' theorem relates a likelihood function $\mathcal{P}(d \vert s)$ to the posterior distribution $\mathcal{P}(s \vert d)$:
\begin{align}
    \mathcal{P}(s \vert d) = \frac{\mathcal{P}(d \vert s)\, \mathcal{P}(s)}{\mathcal{P}(d)} 
    \label{eq:bayes}
\end{align}
Here $\mathcal{P}(s)$ denotes the prior distribution of the parameter $s$, and $\mathcal{P}(d)$ represents the probability of the data $d$, which are usually given already. In IFT, the right side of (\ref{eq:bayes}) is expressed using the \textit{information Hamiltonian} $\mathcal{H}(d, s)=-\ln{(\mathcal{P}(d \vert s)\, \mathcal{P}(s))}$:
\begin{align}
    \mathcal{P}(s \vert d) = \frac{e^{-\mathcal{H}(d, s)}}{\mathcal{Z}(d)}
    \label{eq:posterior_ift}
\end{align}
Here $\mathcal{Z}$ represents the evidence. This reformulation turns probability into an additive quantity.

To formulate a science challenge in this framework, we define the measurement equation:
\begin{align}
    d = R[s] + n
    \label{eq:measurement_equation}
\end{align}
Here $d$ again denotes the data, $s$ the signal parameter of interest, $n$ an additive noise term, and $R$ an operator that maps the signal to the data space. This equation can be linked to the likelihood function. Assuming that the noise $n$ follows a Gaussian statistics $\mathcal{G}(n,N)$ with covariance $N$, the likelihood function reads
\begin{align}
    \mathcal{P}(d\vert s) = \mathcal{G}(d - R[s], N)\, .
\end{align}
This leads us to the \textit{likelihood Hamiltonian}:
\begin{align}
    \mathcal{H}(d\vert s) &= -\ln \mathcal{G}(d - R[s], N) \\
    &= \frac{1}{2}(d - R[s])^\dagger N^{-1} (d - R[s]) + \frac{1}{2}\ln \vert 2 \pi N \vert,
\end{align}

To facilitate corresponding calculations, a frequently used approach is \textit{standardization}, also referred to as \textit{reparametrization trick} \cite{kingma_variational_2015}. 
Standardization means that a coordinate transformation between some \textit{standardized} $\xi$-coordinates and the signal $s$-coordinates is found such that $\xi$ is a priori standard distributed:
\begin{align}
    P(\xi)=P(s)\,\Big|\frac{ds}{d\xi}\Big|=\mathcal{G}(\xi,\mathbbm{1})
\end{align}

These standard coordinates $\xi$ are related to the parameters of interest $s$ via a globally invertible coordinate transformation $s(\xi) = f(\xi)$, that leaves the prior probability measure invariant. Specifically: $P(s)\, ds = P(\xi)\, d\xi = \mathcal{G}(\xi|0, \mathds{1}) \quad \forall s = f(\xi)$. In this frame the standardized Hamiltonian of the joint distribution reads
\begin{align}
    \mathcal{H}(d, \xi) \widehat{=} \mathcal{H}(d\vert f(\xi)) + \frac{1}{2}\xi^\dagger \mathbbm{1}\,\xi,
\end{align}
where terms independent of $\xi$-parameters are dropped. This version of the Hamiltonian will be central to the inference process (see Section~\ref{sec:inference}).
The NIFTy-package \cite{Edenhofer2024} permits us to conveniently implement such a standardized generative forward model and to invert them probablistically using the JAX library \cite{jax2018github}.

\subsection{Python library JAX}

The open source Python library JAX is designed for high-performance numerical computing and large-scale machine learning \cite{jax2018github}. 
For the user it is very similar to the widely-used numpy package~\cite{harris2020array}. Beyond ordinary numpy functionality, JAX offers three key features that make it particularly useful for this inference problem:
\begin{enumerate}
    \item Most JAX code can be automatically differentiated, yielding exact derivatives and gradients. This is referred to as autograd.
    \item Python code can be just-in-time (JIT) compiled under certain conditions, notably writing code in the \textit{functional} style, i.e., no side effects of subroutines. Just-in-time compilation can then adaptively optimize the code and vastly decrease computation times, often over multiple magnitudes.
    \item JAX is able to use NVIDIA's cuda backend~\cite{cuda}, allowing seamless execution on GPUs. This can decrease computation time further.
\end{enumerate}

Next to autograd and JIT compilation, JAX offers the convenience function \verb|vmap|, which maps functions in parallel over arrays. This automatically vectorizes calculations written for a specific task. The function \verb|pmap| is similar, but instead of vectorization, it parallelizes execution over multiple computation cores.

JAX is essential for this inference, as it provides both gradients for the optimization procedure and time-efficient procedures for working with the large arrays required for the shower geometry, the finite-thickness source model, and the retarded-time deposition.

\section{Shower Model \label{showermodel}}

As the shower model we refer to the electric currents in the radio-emitting part of the air shower and to their development as the shower traverses the Earth's atmosphere. The task of inference is to match the observed radio signals with this shower model. To do this, possible current configurations need to be generated and their radio emission, propagation and observation simulated. The shower model is therefore expressed in a way that is suited for a fitting procedure.

We formulate the shower model as a semi-parametric generator
\begin{align}
    G:\xi \mapsto \vec{j},
\end{align}
which maps standardized latent variables $\xi$ to a physically plausible current density $\vec{j}$. All stochastic quantities are represented as deterministic transformations of unit-Gaussian variables. Sampling a shower therefore corresponds to drawing
\begin{align}
    \xi \hookleftarrow \mathcal{G}(0,\mathbbm{1})
\end{align}
and evaluating
\begin{align}
    \vec{j}=G(\xi).
\end{align}
This construction keeps the prior simple in latent space while allowing the generated current density to contain correlated spatial structures, a finite geometrical support, and a finite shower-front thickness.

\paragraph{Correlated current sheet:}

The generator first constructs a vector-valued current sheet over the transverse shower plane and along the discretized shower development using the orthogonal basis (\ref{eq:shower_basis}). 

Three independent latent correlated fields are defined over the shower-development coordinate $q$ and the transverse coordinates $x$ and $y$. In practice, these correlated fields are constructed in the harmonic domain conjugate to the discretized source coordinates $(q,x,y)$, where $q$ labels the shower-development coordinate along $u$. In this representation the translation-invariant covariance is diagonal and specified by component-wise power spectra, and the real-space fields $F_\alpha(q,x,y)$ are obtained by an inverse Fourier transform:
\begin{align}
    F_\alpha
    \hookleftarrow
    \mathrm{CF}\left(0,K_{Q,\alpha}\otimes K_{XY,\alpha}\right),
    \qquad
    \alpha\in\{\mathrm{gm},\mathrm{dip},v\}.
    \label{eq:correlated_fields}
\end{align}
The first argument of $\mathrm{CF}$ fixes the prior mean of each correlated Gaussian field to zero, so that coherent current structures are generated by the latent excitation and the inferred correlation spectra. 
Furthermore, $K_{Q,\alpha}$ describes correlations along the shower development, while $K_{XY,\alpha}$ describes correlations across the transverse plane. 

Thus, the three drawn fields $F_\alpha$ describe the geomagnetic component $\alpha=\mathrm{gm}$, a dipole-like transverse component $\alpha=\mathrm{dip}$, and a component along the shower axis $\alpha=v$. The tensor product $K_{Q,\alpha}\otimes K_{XY,\alpha}$ denotes a separable covariance operator, combining correlations along the shower-development coordinate with correlations across the transverse shower plane.

Before applying the geometrical masks, the current sheet is written as
\begin{align}
    \vec{S}_0(q,x,y)
    =
    \operatorname{softplus}\!\left(F_{\mathrm{gm}}(q,x,y)\right)\vec{e}_{\vec{v}\times\vec{B}}
    -
    \tanh\!\left(\frac{y}{d_{\mathrm{dip}}}\right)
    F_{\mathrm{dip}}(q,x,y)\vec{e}_{\vec{v}\times(\vec{v}\times\vec{B})}
    +
    F_v(q,x,y)\vec{e}_{\vec{v}} .
    \label{eq:current_sheet_unmasked}
\end{align}
The softplus transformation prevents a reversal of the geomagnetic current direction selected by the Lorentz force. The dipole template biases the second transverse component towards an antisymmetric radial structure, while still allowing local variations through the correlated field. The component parallel to the shower axis is left unconstrained.

\paragraph{Geometric masks:}

The correlated fields are approximately homogeneous across their domain, whereas an air shower occupies only a finite longitudinal interval and a finite lateral region around its axis. We therefore apply a scalar mask to all current components,
\begin{align}
    \vec{S}(q,x,y)
    =
    M_{\mathrm{long}}(q)\,
    M_{\perp}(q,r)\,
    \vec{S}_0(q,x,y),
    \qquad
    r=\sqrt{x^2+y^2}.
    \label{eq:masked_current_sheet}
\end{align}

The longitudinal development is described by an unnormalized Gaussian envelope,
\begin{align}
    M_{\mathrm{long}}(q)
    =
    \exp\!\left[
    -\frac{1}{2}
    \left(
    \frac{q-q_c}{\sigma_q}
    \right)^2
    \right],
    \label{eq:longitudinal_mask}
\end{align}
where the center $q_c$ and the width $\sigma_q$ are inferred. This mask localizes the active part of the cascade along the shower axis. Since atmospheric depth depends nonlinearly on its position along the shower trajectory, the same envelope becomes strongly asymmetric when expressed as a function of grammage, qualitatively resembling the longitudinal development of an air shower.

The transverse mask combines a steep radial decrease with a smooth outer boundary,
\begin{align}
    M_{\perp}(q,r)
    =
    \left(1+\frac{r}{r_0}\right)^{-p}
    C\!\left(r;R(q),\Delta r\right).
    \label{eq:transverse_mask}
\end{align}
The parameters $r_0$ and $p$ describe the radial fall-off and are inferred for each shower. The smoothstep function $C$ imposes a smooth cutoff at the outer radius $R(q)$, which is allowed to vary smoothly during the shower development:

\begin{align}
    C(r;R(q),\Delta r) =\begin{cases}
    1, & r \leq R(q) - \Delta r\\
    3t^2 - 2t, & R(q) - \Delta r < r < R(q)\\
    0, &r \geq R(q)
    \end{cases}
\end{align}
where $t = \frac{R(q) - r}{\Delta r}$.
\paragraph{Finite shower-front thickness:}

At this stage, the generator has produced a two-dimensional current sheet for each stage of the shower development. A physical shower front, however, has a finite extension behind its leading surface. We introduce a coordinate $h$, where $h=0$ corresponds to the leading shower front and $h>0$ to positions trailing behind it. The sheet current is redistributed along this coordinate by normalized thickness weights $w_i(r),$ where $i$ labels the sheet behind the front sheet.

The thickness profile is based on the Gamma density
\begin{align}
    g(h;k,\theta)
    =
    \frac{
    h^{k-1}\exp(-h/\theta)
    }{
    \Gamma(k)\theta^k
    },
    \qquad
    h>0 .
    \label{eq:gamma_density}
\end{align}
This composes a smooth, polynomial rise with an exponential decay after some maximum density.
For discrete thickness bins, we use the normalized weights
\begin{align}
    G_i(k,\theta)
    =
    \frac{
    g(|h_i|;k,\theta)\,\Delta h_i
    }{
    \sum_j g(|h_j|;k,\theta)\,\Delta h_j
    } .
    \label{eq:gamma_weights}
\end{align}

The parameters of the main Gamma component vary smoothly with lateral distance. We define
\begin{align}
    \varphi(r)
    =
    \frac{r^2}{r^2+r_{\mathrm{tr}}^2}
    \label{eq:radial_transition}
\end{align}
and use as the parameterizations in (\ref{eq:gamma_density})
\begin{align}
    k(r)
    &=
    k_\varepsilon
    +
    k_{\mathrm{core}}
    +
    k_{\mathrm{gain}}\,
    \varphi(r),\\
    \theta(r)
    &=
    \theta_{\mathrm{core}}
    +
    \theta_{\mathrm{gain}}\,
    \varphi(r).
    \label{eq:gamma_parameters_radial}
\end{align}
Here $r_{\mathrm{tr}}$ is the transition radius and $k_\varepsilon= 0.1$ is a fixed positive offset for numerical stability. The parameters $k_{\mathrm{core}}, k_{\mathrm{gain}}, \theta_{\mathrm{core}},$ and $\theta_{\mathrm{gain}}$ are learnable and follow log-normal priors. Near the shower axis, the profile is controlled by $k_{\mathrm{core}}$ and $\theta_{\mathrm{core}}$. Toward the outer disk, the gain parameters broaden the profile and shift its current farther behind the shower front, thereby describing both increasing thickness and curvature.

An additional central tail component is included to describe a broader trailing population close to the shower axis. Its radial mixing fraction is
\begin{align}
    \lambda(r)
    =
    \lambda_0
    \exp\!\left[
    -\frac{r^2}{2r_{\mathrm{tail}}^2}
    \right].
    \label{eq:tail_mixing}
\end{align}
The final thickness weights are
\begin{align}
    w_i(r)
    =
    \left(1-\lambda(r)\right)
    G_i(k(r),\theta(r))
    +
    \lambda(r)
    G_i(k_{\mathrm{tail}},\theta_{\mathrm{tail}}),
    \qquad
    \sum_i w_i(r)=1 ,
    \label{eq:final_thickness_weights}
\end{align} where $r_{\mathrm{tail}}$ is a fixed parameter estimated from simulated showers.
The normalization ensures that the finite thickness redistributes the current behind the front without changing the integrated current sheet.

\paragraph{From current sheet to volume current density:}

Combining the correlated field, geometric mask, and thickness profile yields a stack of current sheets behind the leading shower front. The index $i$ labels the thickness bin along the $h$-coordinate, while $(q,x,y)$ labels the shower-development coordinate and the position in the transverse shower plane. For each thickness bin $i$, the current sheet is
\begin{align}
    \vec{S}_i(q,x,y)
    =
    M_{\mathrm{long}}(q)\,
    M_{\perp}(q,r)\,
    \vec{S}_0(q,x,y)\,
    w_i(r),
    \label{eq:thick_current_sheet}
\end{align}
and the corresponding piecewise constant volume current density in this layer is
\begin{align}
    \vec{j}_i(q,x,y)
    =
    \frac{\vec{S}_i(q,x,y)}{\Delta h_i}.
    \label{eq:true_currrent}
\end{align}

\paragraph{Timing model:}
As real showers happen at arbitrary times, the generated shower requires a temporal offset. For numerical reasons like tempoeral pixel interpolation and particularly differentiability of the model, this shift is implemented using a blurring kernel.

Let $\tau$ be the shift, we calculate the time shifted electric field $\vec{E}^a(t_k)$ at antenna $a$ and time slice $t_k$ from the unshifted field $\vec{E}^{\,a\prime}$ at unshifted times $t_k^\prime$ as

\begin{align}
    \vec{E}^a(t_k) = \sum_{\eta} w_\tau[\eta] \vec{E}^{\,a\prime}(t_k^\prime - \eta \Delta t),
\end{align}
where $\Delta_t$ is the size of the time binning. The normalized weighting function $w_\tau[\eta]$ is chosen as~\begin{align}
    w_\tau[\eta] = \frac{\exp(\psi(\eta))}{\sum_{\eta^\prime} \exp((\psi(\eta^\prime))},
\end{align}
where $\psi(\eta)$ is a Gaussian random field, and the transformation corresponds to a softmax function. These weights are inferred together with the current and thickness parameters and are shared by all observers.

\paragraph{Shower model parameters:}

Overall, the latent variables of the model can be partitioned into blocks,
\begin{align}
    \xi
    =
    \left(
    \xi_{\mathrm{CF}},
    \xi_{\mathrm{mask}},
    \xi_{\mathrm{height}},
    \xi_{\mathrm{time}}
    \right).
    \label{eq:model_parameters}
\end{align}
The block $\xi_{\mathrm{CF}}$ contains the latent variables of the correlated current fields and their spectra. The block $\xi_{\mathrm{mask}}$ contains the parameters of the longitudinal and transverse masks. The block $\xi_{\mathrm{height}}$ controls the finite shower-front thickness. The block $\xi_{\mathrm{time}}$ contains the global timing weights. These standardized variables are adapted during inference so that the generated currents reproduce the measured antenna traces while remaining consistent with the prior.

\section{Inference\label{sec:inference}}

Our target is to obtain a posterior distribution $\mathcal{P}\left(s\vert d\right)$ (\ref{eq:posterior_ift}) for the particle shower $s$ taking into account the observed data $d$. We follow the measurement equation $d = \mathcal{R}[s] + n$ (\ref{eq:measurement_equation}), which also contains the operator $\mathcal{R}$ for the mapping between the shower $s$ and the observed data $d$ as well as the noise term $n$. 

Our shower model is represented by the standardized model parameters $\xi$ (\ref{eq:model_parameters}). They are transformed by the generator $G$ into the current density $\vec{j}=G(\xi)$, including the correlated current sheet, the geometric masks, the finite shower-front thickness, and the global timing parameters. The response operator $\mathcal{R}$ then maps this generated current density to the electric fields at the antennas. The measurement equation used in the inference is therefore
\begin{align}
    d = \mathcal{R}[G(\xi)] + n .
    \label{eq:full_measurement_equation}
\end{align}

As the inference method, we use the Metric Gaussian Variational Inference (MGVI) \cite{knollmullerMetricGaussianVariational2020,Arras2022}, which is included in NIFTy \cite{Edenhofer2024}. This algorithm is designed for inference in problems with many degrees of freedom and is therefore well suited for the high-dimensional reconstruction of particle showers. In MGVI, the posterior distribution $\mathcal{P}\left(s\vert d\right)$ (\ref{eq:posterior_ift}) is approximated in the standardized latent space by a multivariate Gaussian function:
\begin{align}
 \mathcal{P}\left(s\vert d\right) \equiv \mathcal{P}\left(\xi\vert d\right) \approx \mathcal{G}\left(\xi - \bar{\xi}, \Xi\right)
\label{eq:gauss_mgvi}
\end{align}

Here, the inverse of the Fisher information matrix is used to approximate the covariance matrix $\Xi$. 
The Fisher matrix is a measure of the expected information gain due to the amounts of data.

As the objective function for the inference, we use the Kullback-Leibler divergence. It describes the similarity between the multivariate Gaussian function (\ref{eq:gauss_mgvi}) and the posterior distribution $\mathcal{P}\left(\xi\vert d\right)$ of the shower parameters $\xi$ based on the data $d$:
\begin{align}
D_\mathrm{KL}\left(\mathcal{G}\left(\xi - \bar{\xi}, \Xi\right)\Vert\mathcal{P}\left(\xi\vert d\right)\right)
\end{align}

The posterior distribution $\mathcal{P}\left(\xi\vert d\right)$ can be replaced by the product of likelihood function $\mathcal{P}\left(d \vert \xi \right)$ and the prior distribution $\mathcal{P}\left(\xi \right)$ of the model parameters using Bayes' theorem (\ref{eq:bayes}). 
Since this is a minimization procedure in which the data is already known, the evidence $P(d)$ can be disregarded here.

Using the Hamiltonian language of information field theory, this leads to the following expression, which is used to perform the minimization:
\begin{align}
D_\mathrm{KL}\left(\mathcal{G}\left(\xi - \bar{\xi}, \Xi\right)\Vert\mathcal{P}\left(\xi\vert D\right)\right) 
&=\langle \mathcal{H}(\xi\vert D)\rangle_{\mathcal{G}\left(\xi - \bar{\xi}, \Xi\right)} 
- \langle \mathcal{H}(\xi - \bar{\xi}, \Xi)\rangle_{\mathcal{G}\left(\xi - \bar{\xi}, \Xi\right)}\nonumber\\
&= \langle\mathcal{H}(D\vert\xi)+ \mathcal{H}(\xi)\rangle_{\mathcal{G}\left(\xi - \bar{\xi}, \Xi\right)} 
 - \langle \mathcal{H}(\xi - \bar{\xi}, \Xi)\rangle_{\mathcal{G}\left(\xi - \bar{\xi}, \Xi\right)}
\label{eq:objective_function}
\end{align}
The first term on the right-hand side contains the likelihood function for fitting the standardized model parameters $\xi$ to the data, with the $\xi$ following the Gaussian prior distributions (\ref{eq:gauss_mgvi}). The second term represents the approximation of the shower posterior distribution by the shower model, also generated by the Gaussian distributions (\ref{eq:gauss_mgvi}).

Reconstruction quantities are then accessed by plugging posterior samples of the latent variables $\xi$ into the generator and response functions. This yields posterior samples of the current density, the integrated current sheet, the finite-thickness profile, the timing weights, and the predicted antenna traces. Moments of derived quantities are calculated from these samples, for example the posterior mean field $\langle s\rangle_{(s\vert d)}=\langle f(\xi) \rangle_{(\xi\vert d)}$.

The MGVI algorithm approximates the posterior with a multivariate Gaussian probability density function. This prevents it from accounting for multi-modality, if present. Multi-modality can, however, still be detected by repeating the analysis with varying initial conditions, as those might then end up in different modes. We performed $N_\text{run}$ reconstructions with initial conditions drawn from the prior distribution. We calculate the dispersion of the mean flux images $\overline{I}_i(x)=1/N_\text{samp} \, \sum_j^{N_\text{samp}} I_{ij}(x)$ of the different samples,
\begin{align}
\sigma^2_{\overline{I}} = \frac{1}{N_\text{run}} \sum_i^{N_\text{run}} \int dx^4\,[{\overline{I}}_{i}(x) - {\overline{\overline{I}}}(x)]^2
\end{align}
with $\overline{\overline{I}}(x)=1/N_\text{run} \, \sum_i^{N_\text{run}} \overline{I}_{i}(x)$ the average over all samples of all runs.
This can be compared to the typical dispersion within a run
\begin{align}
\sigma_{I}^2 = \frac{1}{N_{\text{samp}} N_{\text{run}}} \sum_i^{N_{\text{run}}}\sum_j^{N_{\text{samp}}}  \int dx^4\,[{I}_{ij}(x) - \overline{I}_{i}(x)]^2
\end{align}
A ratio of $\sigma_{\overline{I}}^2/ \sigma_{I}^2 \ge 1$ would be an indicator of significant multi-modality.

\section{Performance using Simulated Air Showers}

In this section, we present results obtained using simulated particle showers from the CORSIKA7 library \cite{Heck:1998vt,Heck:1998gr} for air showers from ultra-high-energy cosmic rays. The radio emission from the charged shower particles is generated by the CoREAS library \cite{Huege:2013vt}. Protons or iron nuclei were simulated as primary particles. Their energies range from $2{.}5 \cdot 10^{18}\,\mathrm{eV}$ to $1{.}2 \cdot 10^{20}\,\mathrm{eV}$, and the zenith angle spans the interval from $65^{\circ}$ to $80^{\circ}$. Since precise methods for reconstructing the energy and arrival direction of a cosmic particle already exist, we adopt their values from the simulated truth. As our true reference quantity that is to be inferred, we use the unfiltered electric fields at the antenna positions.

In the following, we first present detailed results from reconstructing a single simulated iron-induced shower. We will subsequently analyse a series of proton and iron showers with regard to the maximum shower development for each.

For the search volume, a cuboid approximately $65$ km long is assumed in the longitudinal direction of the shower. The source is represented on a shower-aligned grid. The longitudinal coordinate follows the shower development, while the associated transverse square area has side lengths of $600\,\meter$ at a resolution of $25\times 25$ pixels. In addition, each transverse source element is extended along the thickness coordinate $h$ behind the leading shower front, which is discretized into a finite number of layers. The normalized weights along this coordinate are generated by the radial thickness model and enter the retarded-time deposition in the forward calculation. The inference time on a GPU (NVIDIA RTX 6000 Ada Generation) was roughly $3\,\si{\hour}$ for full posterior sampling, and roughly $20\,\si{\minute}$ for a MAP estimate.

\subsection{Benchmarks of an iron-induced air shower}

\begin{figure}[h!]
    \centering
    \includegraphics[width=0.4\linewidth]{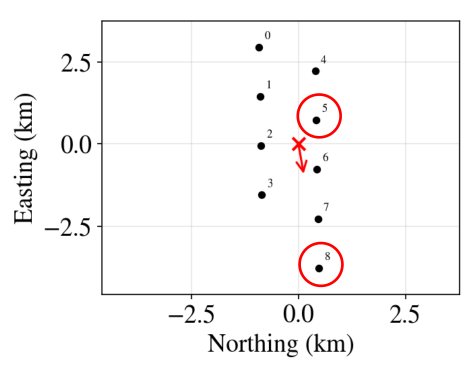}
    \caption{Antenna layout for one iron-induced air shower reconstructed in this section. The numbers denote the observer designations used. The star and arrow denote the impact and shower axis. The circles indicate two observers with detailed investigations below.}
    \label{fig:antenna_field}
\end{figure}

The iron-induced shower shows energy of $3{.}1 \cdot 10^{18}\,\mathrm{eV}$, a zenith angle of $69.3^{\circ}$ and an azimuth angle of $97.7^{\circ}$. The antenna positions with signals are shown in Figure~\ref{fig:antenna_field} with the star symbol denoting the shower center of gravity and the arrow indicating the shower direction. In addition, two antennas are marked which are referred to with further detailed investigations.

\paragraph{Electric field reconstruction at the antenna:}

\begin{figure}[h!]
    \centering
    \includegraphics[width=0.9\linewidth]{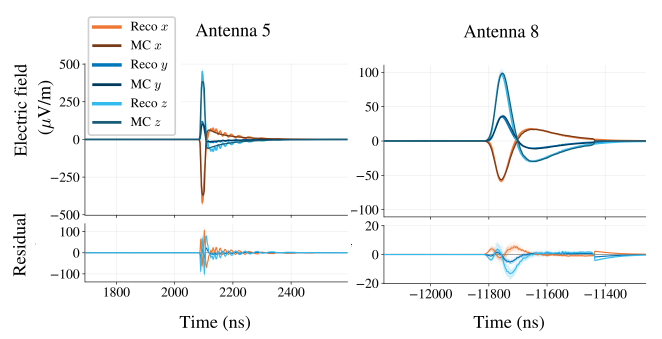}
    \caption{Simulated and reconstructed electric field for observers of one iron-induced air shower at the positions $5$ and $8$. The top panel shows the Monte Carlo truth and the individual posterior samples. The bottom panel shows the residual, with a shaded $1\,\sigma$-region.}
    \label{fig:antenna_efield}
\end{figure}

The simulated electric fields at the antenna positions represent the target quantities that the generative model must reproduce from the varying currents in the particle shower and their corresponding electric fields at the antennas. Figure~\ref{fig:antenna_efield} shows the simulated electric fields and those reconstructed by the generative model for the two observer positions $5$ and $8$, together with their residuals. Overall, the reconstruction accurately reproduces the time histories, with good agreement in terms of signal timing, duration and amplitude.

Antenna position $5$ is located near the Cherenkov ring, where a ringing artefact becomes visible. Here, the rapid time compression of the signal makes the trace sensitive to the finite discretisation of the shower-front thickness in the generative model and in the retarded-time renderer. This effect can be reduced by using a higher numerical resolution in the thickness coordinate and in the source discretisation, at the cost of increased computational effort.

\paragraph{Vector field of currents in the shower disc:}

\begin{figure}[h!]
    \centering
    \includegraphics[width=\linewidth]{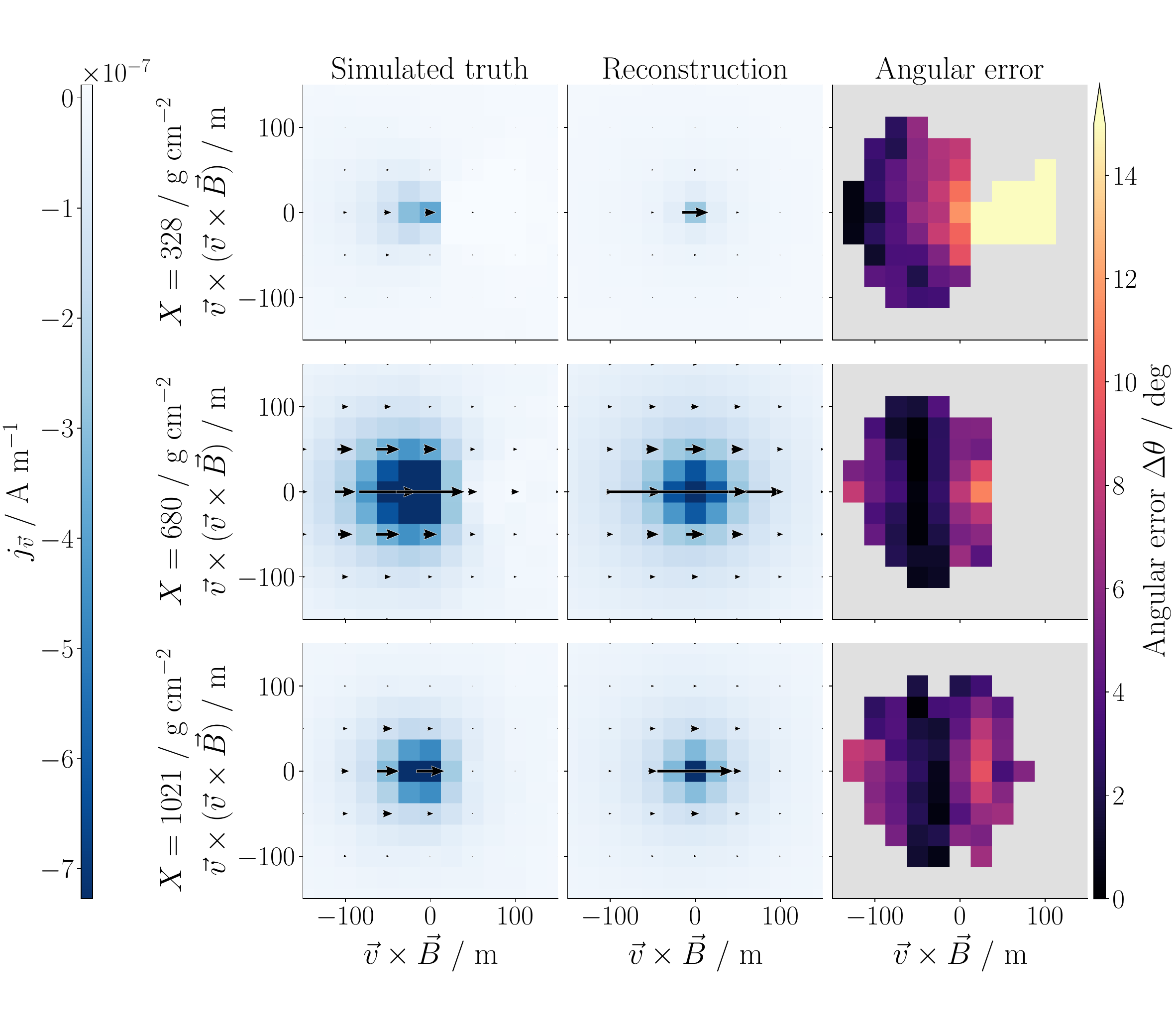}
    \caption{Excerpt of the shower sheet with projected current vectors $\vec{j}$ at different stages of one iron-induced air shower. The arrows indicate the current directions transverse to the shower development, projected onto the geomagnetic emission direction $\vec{e}_{\vec{v}\times\vec{B}}$. The color indicates the longitudinal current component following the shower axis $\vec{e}_{\vec{v}}$. Left: Monte Carlo generated projected current densities. Center: Reconstructed projected current densities. Right: Angular uncertainty in the reconstruction of the projected current densities. }
    \label{fig:antenna_current_vectors}
\end{figure}

Figure~\ref{fig:antenna_current_vectors} compares the Monte Carlo and reconstructed current vectors after projection onto the plane spanned by the shower-axis direction $\vec{e}_{\vec{v}}$ and the geomagnetic direction $\vec{e}_{\vec{v}\times\vec{B}}$.

In the left-hand column, the projected Monte Carlo current vectors $\vec{j}^{\textrm{MC}}$ near the shower axis are shown for three different stages of shower development, expressed in terms of the atmospheric depth $X$. The transverse component is shown along the unit vectors $\vec{e}_{\vec{v}\times\vec{B}}$ along the geomagnetic emission, whilst the longitudinal component $\vec{e}_{\vec{v}}$ follows the shower direction and is shown in colour-coded form.

The second column shows the reconstructed projected current vectors $\vec{j}^{\textrm{rec}}$. It can be seen that their longitudinal component $\vec{j}^{\textrm{rec}}\cdot\vec{e}_{\vec{v}}$ and the component related to the geomagnetic emission $\vec{j}^{\textrm{rec}}\cdot\vec{e}_{\vec{v}\times\vec{B}}$ are reasonably reproduced near the shower axis. This can be seen in particular in the right-hand column, which shows the angular deviation $\Delta\vartheta$ between the true and reconstructed projected current vectors, which lie below $\Delta\vartheta\approx 10^\circ$. The corresponding magnitudes are estimated here with an accuracy of approximately $\pm 50\%$.

For the first time, this gives us access to detailed information within the shower disk. The macroscopic currents reflect local motions of the electrically charged particles in the disk. At least two important vector components of these currents can now be tracked throughout the entire shower development.

The generator includes in addition the dipole-like transverse component along $\vec{e}_{\vec{v}\times(\vec{v}\times\vec{B})}$ for physical completeness and to allow radial current structures in the shower plane. However, this component is only weakly constrained by the antenna signals in the present geometry. The Askaryan emission contributes in the transverse plane in a direction radial to the shower axis, and these contributions largely cancel each other for distant observers. We therefore focus in the following on the robustly constrained longitudinal and transverse geomagnetically induced components.

\paragraph{Shape of the shower disc:}

The finite shower-front thickness is not reconstructed as an independent current value in every four-dimensional voxel. Instead, it is represented by the radial thickness model of the generator. Posterior samples of the latent thickness parameters define the normalized profile $w_i(r)$ along the coordinate $h$. Since this profile is normalized, it redistributes the current behind the leading front without changing the integrated current sheet.

\begin{figure}[h!]
    \centering
    \includegraphics[width=0.5\linewidth]{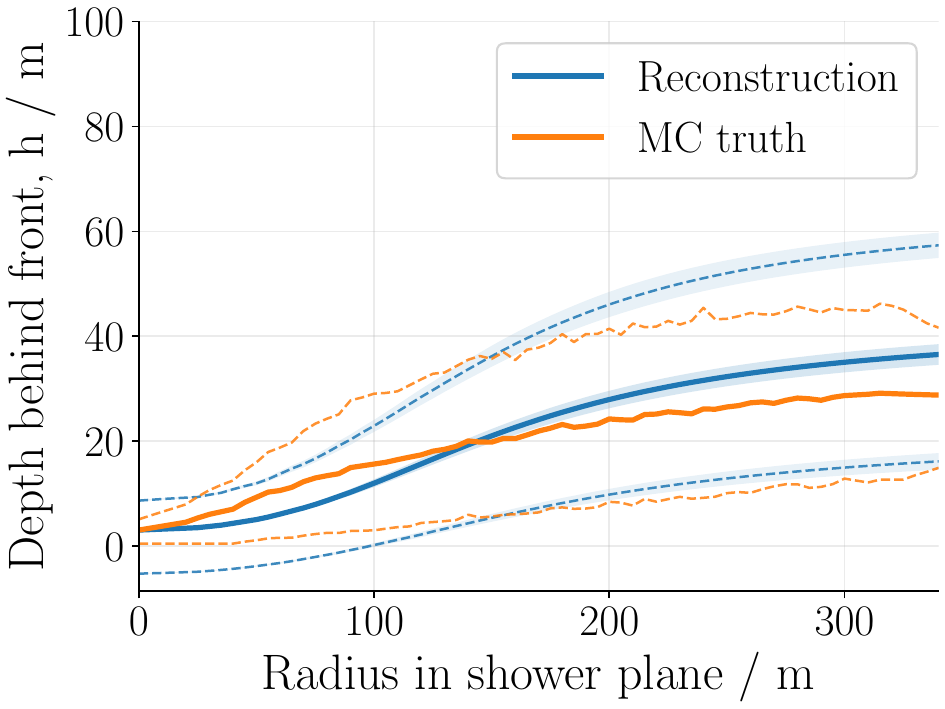}
    \caption{Reconstructed and true thickness for one iron-induced air shower. The full curves refer to the median of the current densities, and the dashed curves to the $16\%$ and $84\%$ quantiles.}
    \label{fig:shower_radius}
\end{figure}

Figure~\ref{fig:shower_radius} shows the effective thickness of the shower disc. The solid lines denote the median of the current-density profile, while the dashed lines represent the $16\%$ and $84\%$ quantiles. The thickness is about constant in time for the reconstructed shower. The Monte Carlo thickness has been evaluated at the true atmospheric depth of the shower maximum $X_{\textrm{max}}$. The reconstruction reproduces the broad radial dependence of the Monte Carlo reference, including the increase at larger lateral distances.

\paragraph{Longitudinal evolution of the integrated current components:}

\begin{figure}[h!]
    \centering
    \includegraphics[width=0.5\linewidth]{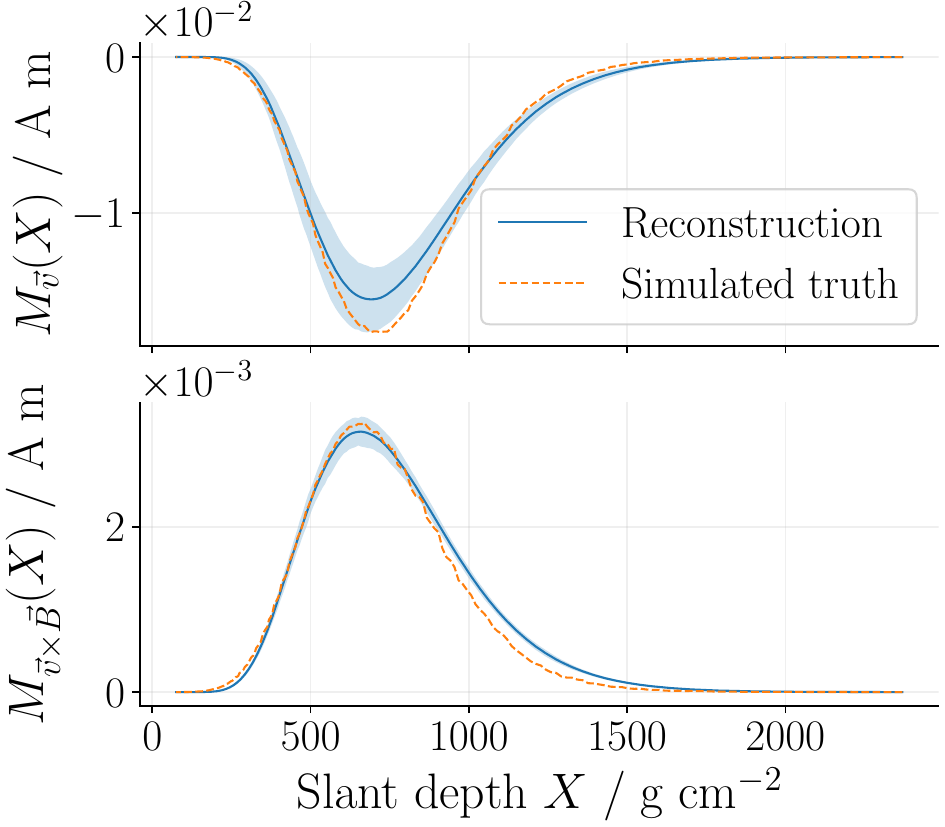}
    \caption{Longitudinal development of the individual current polarizations, integrated over the shower plane of one iron-induced air shower. The reconstruction recovers the distinct longitudinal development of the longitudinal and geomagnetic-transverse current components.}
    \label{fig:long_projections}
\end{figure}

Figure~\ref{fig:long_projections} shows the longitudinal evolution of the longitudinal current component $\vec{j}^{\textrm{rec}}\cdot\vec{e}_{\vec{v}}$ and the geomagnetic-transverse component $\vec{j}^{\textrm{rec}}\cdot\vec{e}_{\vec{v}\times\vec{B}}$. For example, for a polarization component $\ell$, we integrate the reconstructed current density over the transverse shower plane and over the finite thickness of the shower front,
\begin{align}
  M_\ell(X)
  &=
  \int dA \int dh\,
  \vec{j}(X,x,y,h)\cdot\vec{e}_\ell \nonumber\\
  &=
  \int dA\,
  \vec{S}(X,x,y)\cdot\vec{e}_\ell .
  \label{eq:integrated_current_components}
\end{align}

Throughout the entire development of the shower, both of the reconstructed current components correspond very closely to the true simulated current integrals.

\subsection{Reconstruction of the air shower maximum}

\begin{figure}[h!]
    \centering
    \includegraphics[width=0.55\linewidth]{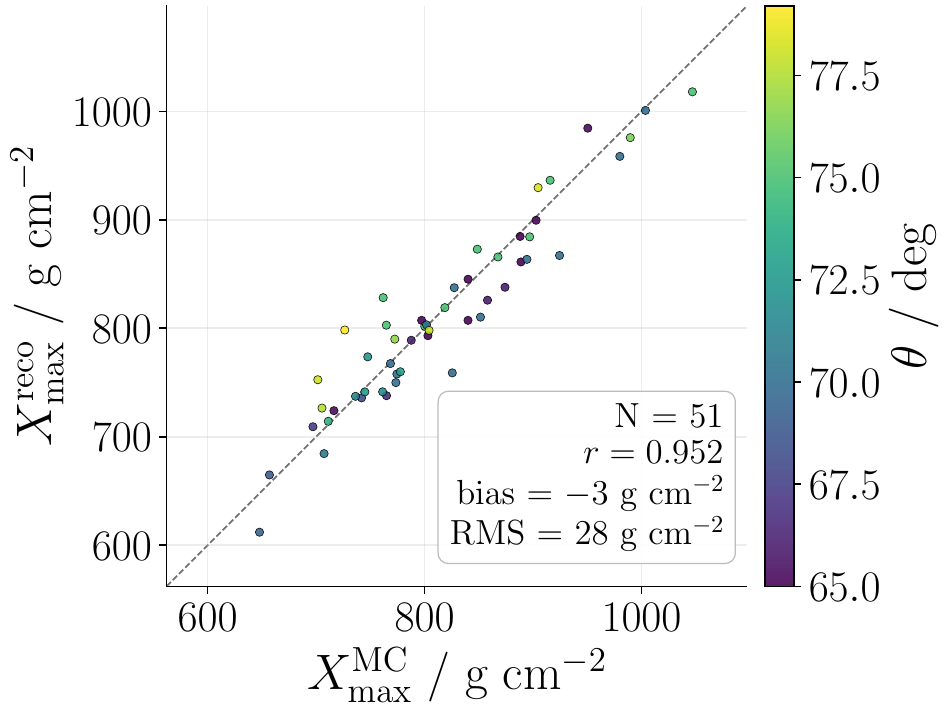}    
    \caption{Reconstructed $X_{\textrm{max}}^{\textrm{rec}}$ from the maximum integrated current density compared with the Monte Carlo $X_{\textrm{max}}^{\textrm{MC}}$ obtained from the Gaisser-Hillas fit performed by CORSIKA for $51$ proton- and iron-induced air showers with energies ranging from $2{.}5 \cdot 10^{18}\,\mathrm{eV}$ to $1{.}2 \cdot 10^{20}\,\mathrm{eV}$. The color indicates the zenith angle of the shower.}
    \label{fig:xmax_reconstruction}
\end{figure}

Here we present the reconstruction of the maximum of the shower development at the atmospheric depth $X_{\textrm{max}}$. To do this, we use the maximum a posteriori estimate of the largest integrated current density during the shower development. Figure~\ref{fig:xmax_reconstruction} shows the reconstructed shower maximum $X_{\textrm{max}}^{\textrm{rec}}$ as a function of the true simulated shower maximum $X_{\textrm{max}}^{\textrm{MC}}$. In the simulation, the shower maximum is determined within the CORSIKA programme by a Gaisser–Hillas fit to the particle density. A total of $51$ proton- and iron-induced showers are shown, which were reconstructed very precisely regardless of their energy and zenith angle.

The high quality is reflected in a Pearson correlation coefficient of $\rho=0.95$, a bias of just $-3 \,\mathrm{g\,cm^{-2}}$ and a root mean square (RMS) deviation of $28 \,\mathrm{g\,cm^{-2}}$. This demonstrates that conventional shower observables such as $X_{\mathrm{max}}$ can be derived from the reconstructed current distribution, rather than being introduced as separate reconstruction parameters.

\subsection{Integrated current and primary energy}
The total current inside an air shower is related to the number of electromagnetic particles and consequently to the calorimetric energy deposited by the shower. Since the calorimetric energy is closely related to the primary particle energy, the integrated current is expected to provide a proxy for the latter.
We define the total integrated current as
\begin{align}
  I
  &=
  \int dX \int dA \left|\int dh\,
  \vec{j}(X,x,y,h)\right|
 .
    \label{eq:total_current}
\end{align}
\begin{figure}[h!]
    \centering
    \includegraphics[width=0.55\linewidth]{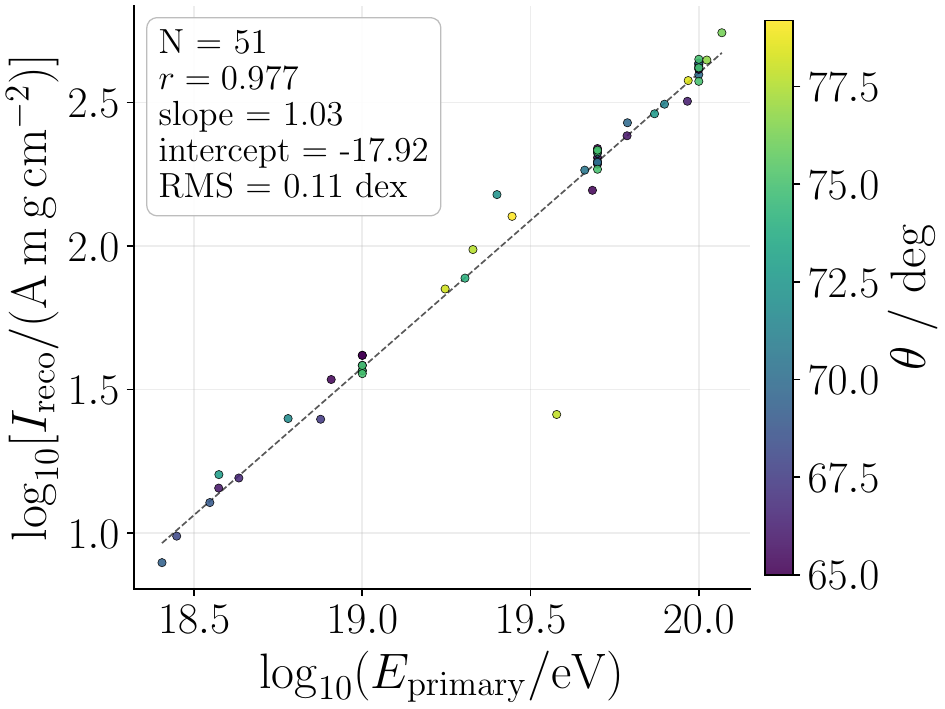}    
    \caption{Integrated current magnitude compared with the Monte Carlo primary energy. Shown are $51$ proton- and iron-induced air showers with energies ranging from $2{.}5 \cdot 10^{18}\,\mathrm{eV}$ to $1{.}2 \cdot 10^{20}\,\mathrm{eV}$. The color indicates the zenith angle of the shower.}
    \label{fig:energy_reconstruction}
\end{figure}
Figure~\ref{fig:energy_reconstruction} shows the integrated current as a function of primary energy on logarithmic axes. Across the $51$ reconstructed showers, the two quantities exhibit a strong correlation in log-space, with a Pearson coefficient of $r = 0.98.$ This indicates that the reconstruction preserves the energy dependence of the overall current magnitude, despite the large number of degrees of freedom in the inferred current distribution. Notably, energy is not an explicit parameter of the reconstruction. The remaining scatter may arise from both shower-to-shower variations and reconstruction uncertainties, which would need to be characterized when calibrating an energy estimator.

One prominent outlier exhibits a substantially underestimated integrated current. In this event, the longitudinal component $j_{\vec{v}}$ is strongly suppressed in the reconstruction, likely due to unfavorable observer geometry providing insufficient constraints on this component. Despite this suppression, the reconstructed $X_{\mathrm{max}}$ remains close to the Monte Carlo value, indicating that the maximum can still be recovered when the integrated current magnitude is underestimated.
\section{Summary}

We have developed a novel Bayesian imaging algorithm for large-scale air showers. Utilizing information field theory (IFT) and its variational inference together with the computational framework JAX, our method simultaneously adjusts numerous latent parameters to reconstruct macroscopic current distributions in the atmosphere from measured electric fields. The source model combines flexible correlated current fields in the shower plane with physically motivated longitudinal and lateral masks, a finite radial shower-front thickness profile, and a global timing model.

The finite-thickness component redistributes the current behind the leading shower front and is propagated through a differentiable retarded-time renderer. This allows the model to reproduce the temporal structure of the measured electric-field traces, in particular for signals affected by Cherenkov compression. The inference algorithm computes approximate posterior distributions, providing statistical insights into the reconstructed currents and derived shower quantities.

Using simulated air showers, our approach demonstrates that the distributions of the two robustly constrained current components within the shower plane, namely the longitudinal and geomagnetically transverse current vector components, can be reconstructed. The method also provides access to derived quantities such as the effective shower-front thickness, the atmospheric depth of the shower maximum, and the primary particle's energy. This enables a detailed characterization of the electromagnetic component of particle showers in terms of their spatial extent and temporal evolution.

The new access to longitudinal and transverse details of shower evolution may provide a basis for future studies of air-shower physics beyond conventional scalar observables. This is particularly interesting for comparisons between showers initiated by different primary particles, such as protons and heavy nuclei. Since heavy nuclei can generate more complex sub-shower structures, a sufficiently precise reconstruction of the current distribution could eventually improve event-by-event sensitivity to the primary mass. Establishing such a connection, however, requires dedicated studies with larger simulated data sets and refined uncertainty quantification.

\acknowledgments
We wish to thank Olaf Scholten, Philipp Windischhofer and Tim Huege for most valuable discussions. This work is supported by the Ministry of Innovation, Science, and Research of the State of North Rhine-Westphalia, as well as by the Federal Ministry of Research, Technology, and Space (BMFTR) in Germany, under their ErUM-IFT funded project, designated for PF as 05D23E01 and for MS as 05D2022. TE furthermore gratefully acknowledge the financial support provided by the German Federal Ministry of Research, Technology and Space (BMFTR) in the framework of the Knowledge creates perspectives for the region!, for the project StStG -- DZA -- Aufbauphase: Deutsches Zentrum f\"ur Astrophysik, Gro{\ss}forschungszentrum in der s\"achsischen Lausitz: Aufbauphase 2026, grant number 03WSP1746. Language editing and text-polishing support was received from DeepL and OpenAI language models.


\begin{thebibliography}{99}

\bibitem{jackson_classical_1999}
J.D.~Jackson, \emph{Classical electrodynamics}, Wiley, New York, {NY}, 3rd ed.~ed. (1999).

\bibitem{Escudie:2019tlt}
A.~Escudie, D.~Charrier, R.~Dallier, D.~Garc\'\i{}a-Fern\'andez, A.~Lecacheux, L.~Martin et~al., \emph{{Radio detection of atmospheric air showers of particles}},  \href{https://arxiv.org/abs/1903.02889}{{arXiv:1903.02889}}.

\bibitem{LOPES:2021ipp}
{\scshape LOPES} collaboration, \emph{{Final results of the LOPES radio interferometer for cosmic-ray air showers}}, \href{https://doi.org/10.1140/epjc/s10052-021-08912-4}{\emph{Eur. Phys. J. C} {\bfseries 81} (2021) 176} [\href{https://arxiv.org/abs/2102.03928}{{arXiv:2102.03928}}].

\bibitem{PierreAuger:2012ker}
{\scshape Pierre Auger} collaboration, \emph{{Antennas for the Detection of Radio Emission Pulses from Cosmic-Ray}}, \href{https://doi.org/10.1088/1748-0221/7/10/P10011}{\emph{JINST} {\bfseries 7} (2012) P10011} [\href{https://arxiv.org/abs/1209.3840}{{arXiv:1209.3840}}].

\bibitem{PierreAuger:2016vya}
{\scshape Pierre Auger} collaboration, \emph{{Measurement of the Radiation Energy in the Radio Signal of Extensive Air Showers as a Universal Estimator of Cosmic-Ray Energy}}, \href{https://doi.org/10.1103/PhysRevLett.116.241101}{\emph{Phys. Rev. Lett.} {\bfseries 116} (2016) 241101} [\href{https://arxiv.org/abs/1605.02564}{{arXiv:1605.02564}}].

\bibitem{PierreAuger:2023lkx}
{\scshape Pierre Auger} collaboration, \emph{{Demonstrating Agreement between Radio and Fluorescence Measurements of the Depth of Maximum of Extensive Air Showers at the Pierre Auger Observatory}}, \href{https://doi.org/10.1103/PhysRevLett.132.021001}{\emph{Phys. Rev. Lett.} {\bfseries 132} (2024) 021001} [\href{https://arxiv.org/abs/2310.19963}{{arXiv:2310.19963}}].

\bibitem{Thoudam:2015lba}
S.~Thoudam et~al., \emph{{Measurement of the cosmic-ray energy spectrum above 10$^{16}$ eV with the LOFAR Radboud Air Shower Array}}, \href{https://doi.org/10.1016/j.astropartphys.2015.06.005}{\emph{Astropart. Phys.} {\bfseries 73} (2016) 34} [\href{https://arxiv.org/abs/1506.09134}{{arXiv:1506.09134}}].

\bibitem{wernerMacroscopicTreatmentRadio2008}
K.~Werner and O.~Scholten, \emph{Macroscopic {{Treatment}} of {{Radio Emission}} from {{Cosmic Ray Air Showers}} based on {{Shower Simulations}}}, \href{https://doi.org/10.1016/j.astropartphys.2008.04.004}{\emph{Astroparticle Physics} {\bfseries 29} (2008) 393} [\href{https://arxiv.org/abs/0712.2517}{{arXiv:0712.2517}}].

\bibitem{Alvarez-Muniz:2014wna}
J.~Alvarez-Mu\~niz, W.R.~Carvalho, Jr., H.~Schoorlemmer and E.~Zas, \emph{{Radio pulses from ultra-high energy atmospheric showers as the superposition of Askaryan and geomagnetic mechanisms}}, \href{https://doi.org/10.1016/j.astropartphys.2014.04.004}{\emph{Astropart. Phys.} {\bfseries 59} (2014) 29} [\href{https://arxiv.org/abs/1402.3504}{{arXiv:1402.3504}}].

\bibitem{Alameddine:2024cyd}
J.M.~Alameddine et~al., \emph{{Simulating radio emission from particle cascades with CORSIKA 8}}, \href{https://doi.org/10.1016/j.astropartphys.2024.103072}{\emph{Astropart. Phys.} {\bfseries 166} (2025) 103072} [\href{https://arxiv.org/abs/2409.15999}{{arXiv:2409.15999}}].

\bibitem{Edenhofer2024}
G.~Edenhofer, P.~Frank, J.~Roth, R.H.~Leike, M.~Guerdi, L.I.~Scheel-Platz et~al., \emph{Re-envisioning numerical information field theory (nifty.re): A library for gaussian processes and variational inference}, \href{https://doi.org/10.21105/joss.06593}{\emph{Journal of Open Source Software} {\bfseries 9} (2024) 6593} [\href{https://arxiv.org/abs/2402.16683}{{arXiv:2402.16683}}].

\bibitem{ensslinInformationTheoryFields2019}
T.A.~En{\ss}lin, \emph{Information theory for fields}, \href{https://doi.org/10.1002/andp.201800127}{\emph{Annalen der Physik} {\bfseries 531} (2019) 1800127} [\href{https://arxiv.org/abs/1804.03350}{{arXiv:1804.03350}}].

\bibitem{jax2018github}
J.~Bradbury, R.~Frostig, P.~Hawkins, M.J.~Johnson, C.~Leary, D.~Maclaurin et~al., \emph{{JAX}: composable transformations of {P}ython+{N}um{P}y programs}, \url{http://github.com/jax-ml/jax}, 2018.

\bibitem{Riegler:2020tzx}
W.~Riegler and P.~Windischhofer, \emph{{Signals induced on electrodes by moving charges, a general theorem for Maxwell\textquoteright{}s equations based on Lorentz-reciprocity}}, \href{https://doi.org/10.1016/j.nima.2020.164471}{\emph{Nucl. Instrum. Meth. A} {\bfseries 980} (2020) 164471} [\href{https://arxiv.org/abs/2001.10592}{{arXiv:2001.10592}}].

\bibitem{kraus_1988}
J.D.~Kraus, \emph{Antennas}, McGraw-Hill, New York, {NY}, 2nd ed.~ed. (1988).

\bibitem{straub2026}
M.~Straub, \emph{Bayesian Imaging of Radio-Emitting Currents Inside Cosmic-Ray Air Showers}, PhD thesis, RWTH Aachen University, 2026

\bibitem{correlatedfield}
P.~{Arras}, H.L.~{Bester}, R.A.~{Perley}, R.~{Leike}, O.~{Smirnov}, R.~{Westermann} et~al., \emph{{Comparison of classical and Bayesian imaging in radio interferometry. Cygnus A with CLEAN and resolve}}, \href{https://doi.org/10.1051/0004-6361/202039258}{\emph{Astronomy and Astrophysics} {\bfseries 646} (2021) A84}.

\bibitem{kingma_variational_2015}
D.P.~Kingma, T.~Salimans and M.~Welling, \emph{Variational dropout and the local reparameterization trick},  \href{https://arxiv.org/abs/1506.02557}{{arXiv:1506.02557}}.

\bibitem{harris2020array}
C.R.~Harris, K.J.~Millman, S.J.~van~der Walt, R.~Gommers, P.~Virtanen, D.~Cournapeau et~al., \emph{Array programming with {NumPy}}, \href{https://doi.org/10.1038/s41586-020-2649-2}{\emph{Nature} {\bfseries 585} (2020) 357}.

\bibitem{cuda}
NVIDIA, P.~Vingelmann and F.H.~Fitzek, \emph{Cuda, release: 10.2.89}, \url{https://developer.nvidia.com/cuda-toolkit}, 2020.

\bibitem{knollmullerMetricGaussianVariational2020}
J.~Knollm{\"u}ller and T.A.~En{\ss}lin, \emph{Metric {{Gaussian Variational Inference}}},  \href{https://arxiv.org/abs/1901.11033}{{arXiv:1901.11033}}.

\bibitem{Arras2022}
P.~Arras, P.~Frank, P.~Haim, J.~Knollm{\"u}ller, R.~Leike, M.~Reinecke et~al., \emph{Variable structures in M87* from space, time and frequency resolved interferometry}, \href{https://doi.org/10.1038/s41550-021-01548-0}{\emph{Nature Astronomy} {\bfseries 6} (2022) 259}.


\bibitem{Heck:1998vt}
D.~Heck, J.~Knapp, J.~N.~Capdevielle, G.~Schatz and T.~Thouw, \emph{CORSIKA: A Monte Carlo code to simulate extensive air showers}, FZKA-6019.

\bibitem{Heck:1998gr}
D.~Heck and J.~Knapp, \emph{Upgrade of the Monte Carlo code CORSIKA to simulate extensive air showers with energies {\ensuremath{>}} $10^{20}-eV$}, FZKA-6097.

\bibitem{Huege:2013vt}
T.~Huege, M.~Ludwig and C.~W.~James,
\emph{Simulating radio emission from air showers with CoREAS}, AIP Conf. Proc. \textbf{1535} (2013) no.1, 128, doi:10.1063/1.4807534, \href{https://arxiv.org/abs/1301.2132}{arXiv:1301.2132 [astro-ph.HE]}


\end{thebibliography}
\end{document}